\documentclass[prl,twocolumn]{revtex4-1}

\usepackage{amsmath}
\usepackage{amssymb}
\usepackage{amsthm}

\usepackage{graphicx}
\usepackage{graphics}
\usepackage{bbold,bm}
\usepackage{hyperref}
\usepackage{epsfig}
\usepackage{dcolumn}

\newcommand\vex[1]{\mathbf{#1}}
\newcommand\gvex[1]{\boldsymbol{#1}}

\def\sgn{\mathrm{sgn}}

\def\tr{\mathrm{Tr}}

\def\id{\mathbb{1}}

\def\dbar{\hbox{$d$\kern-0.6em\raise0.3em\hbox{$-$}}\hspace{-0.5mm}}

\begin{document}

\title{Zero modes of the generalized fermion-vortex system in magnetic field}

\author{Chi-Ken~Lu} 
\address{Department of Physics, Indiana University, Bloomington, Indiana 47405, USA}

\author{Babak~Seradjeh}
\address{Department of Physics, Indiana University, Bloomington, Indiana 47405, USA}

\begin{abstract}
We show that Dirac fermions moving in two spatial dimensions with a generalized dispersion $E\sim p^N$, subject to an external magnetic field and coupled to a complex scalar field carrying a vortex defect with winding number $Q$ acquire $NQ$ zero modes. This is the same as in the absence of the magnetic field. Our proof is based on selection rules in the Landau level basis that dictate the existence and the number of the zero modes. We show that the result is insensitive to the choice of geometry and is naturally extended to general field profiles, where we also derive a generalization of the Aharonov-Casher theorem. Experimental consequences of our results are briefly discussed.
\end{abstract}

\maketitle

\emph{Introduction}.---%
Topological defects such as kinks in one spatial dimension and vortices and domain walls in higher dimensions frequently arise in physical systems as excitations in a background quantum field or an ordered state of matter. It was understood long ago that fermions moving in the background of a vortex defect can acquire  fractional quantum numbers relative to the smooth vacuum background~\cite{JacReb76a,SuSchHee79a,JacSch81a,NieSem83a}. This fractionalization is mediated by zero-energy bound states of fermions to the vortex. At a domain wall, bound states turn into mid-gap propagating modes along the wall that endow it with special transport properties~\cite{SemSemZho08a}. In either case the mid-gap spectrum and, in particular, the number of zero modes are central to understanding the physical properties of the combined system of fermions and defects~\cite{LuHer12a}. Formally, the number of zero-energy states is given by the index of the Dirac Hamiltonian of the fermion-vortex system.  When the spectrum is symmetric, i.e. there is an operator $\Gamma$ that anticommutes with the Dirac Hamiltonian $H$, we may define the index~\cite{Note1}
\begin{equation}\label{eq:ind}
  \mathrm{ind}\,H \equiv \tr(\Gamma)=\nu_+-\nu_-,
\end{equation}
where $\nu_\pm$ is the number of zero modes that are eigenstates of $\Gamma$ with eigenvalue $\pm1$. Index theorems relate this number to the topological information encoded in the scalar field~\cite{AtiSin63a}.

An important system where such configurations arise is the single- or multi-layer graphene with a spectral gap due to a U(1) order parameter, such as Kekul\'e bond density~\cite{HouChaMud07a,WeeFra10a}, antiferromagnetic~\cite{Her06b,Kha12a,ZhaMinMac12a}, superconducting~\cite{Her10b,GhaWil12a,HosZar12a,MurVaf13a}, and quantum anomalous Hall~\cite{LuHer12a} states. The effective Hamiltonian of fermions in a system of $N$ Bernal-stacked layers of graphene is a generalized Dirac form where the spinor structure comes from the sublattices of the honeycomb lattice and the spectrum has band touchings with energy $E\sim p^N$. Similar structures may also arise by bringing together two graphene films separated by a dielectric spacer~\cite{LozSok08a,MinBisSu08b,ZhaJog08b,SerWebFra08a,Ser12b,PerNeiHam13a}. Then the order parameter describes the superfluid state of interlayer excitons driven by the Coulomb interaction between the electrons and holes on the two films. For $N=1$, the zero modes
are equivalent to those first found by Jackiw and Rossi~\cite{JacRos81a} and later studied in condensed-matter systems~\cite{HouChaMud07a,Her07a,Ser08a,SerWeeFra08a,SerMooFra09a}. The index is given by the total vorticity~\cite{Wei81a}.  This result was later extended to the case with chiral and regular orbital fields~\cite{Her10a,HouChaMud10a}. The mid-gap spectrum in this system has been argued to explain the critical behavior observed in graphene at high fields~\cite{CheLiOng09a}.  However, previous work used special forms of the Hamiltonian or arguments that seem difficult to extend to $N>1$, vortices of winding number $Q>1$, and general field profiles.

In this paper, we focus on the structure of the Hamiltonian and its selection rules instead of the details of the corresponding differential equation to study the generalized fermion-vortex problem in magnetic field. The selection rules can be most easily seen in the Landau level basis when the order parameter vanishes. Working with a uniform magnetic field first we show that selection rules reflect conserved quantities depending on the choice of geometry and gauge. For example, in the ribbon geometry and Landau gauge they reflect conservation of transverse momentum, while in the Corbino disk geometry and symmetric gauge they reflect the conservation of angular momentum. When the order parameter is adiabatically turned on, the selection rules ensure the index in Eq.~(\ref{eq:ind}) is exactly $NQ$. For a general field profile, Landau levels are mixed. Nevertheless, the zeroth Landau level persists, a result first discovered by Aharonov and Casher~\cite{AhaCas79a} in the case $N=1$. We extend this result to $N>1$ and show that the selection rules in this case also yield $NQ$ zero-energy states so long as the gap is not closed. This is the same number obtained in the absence of magnetic field. Thus, we show that the magnetic field does not affect the index.
 
\emph{Continuum hamiltonian}.---%
Our starting point is the continuum Hamiltonian,
\begin{equation}
  H = \gamma_0\gamma_1 V_1+ \gamma_0\gamma_2 V_2 + M.
  \label{eq:Ham}
\end{equation}
The kinetic energy in the first term is given in terms of the operators 
\begin{equation}
V_1+i V_2 = \alpha (p_x+i p_y)^N,
\end{equation}
where $(p_x,p_y)\equiv\vex p =-i\gvex\nabla$ is the momentum operator, $\alpha$ is a parameter, and $N$ is a positive integer. The matrices $\gamma_0$ and $\gvex\gamma=(\gamma_1,\gamma_2,\gamma_3)$ form a Clifford algebra $\{\gamma_\mu,\gamma_\nu\}=2\delta_{\mu\nu}$. For concreteness, we use the Weyl representation $\gamma_0=\sigma_1\otimes\id, \gvex\gamma = -i\sigma_2\otimes\gvex\sigma$, and $\gamma_5 = i\gamma_0\gamma_1\gamma_2\gamma_3 = \sigma_3\otimes\id$, where $\gvex\sigma$ are the Pauli matrices. The second term specifies the order parameter, $M=M_1 \gamma_0 + i M_2 \gamma_0\gamma_5$ with $M_1+iM_2=|M(\vex r)|e^{-i\chi(\vex r)}$. When $N=1$, i.e. in the single-layer graphene $\alpha$ is the Fermi velocity and the Hamiltonian acts on the spinor $\psi=(\psi_{\mathrm{A}+},\psi_{\mathrm{B}+},\psi_{\mathrm{B}-}, \psi_{\mathrm{A}-})^\intercal$, where A and B are the two sublattices and the $\pm$ represent the valley index. 
In the $N$-layer graphene this Hamiltonian provides a low-energy effective description of the system with reduced degrees of freedom~\cite{McCFal06a}.

The magnetic field is included through the substitution $\vex p \mapsto \vex p-\vex A$, where $\vex A$ is the vector potential. Then $H=\sigma_z\otimes h(\vex A) + M$, where 
$
h(\vex A)=\alpha \left( \begin{array}{cc} 0 & {\pi^\dagger}^N  \\ \pi^N  & 0 \end{array} \right)
$,
and $\pi = p_x-A_x+i(p_y - A_y)$ satisfies $[\pi,\pi^\dagger]=2B$. For a uniform magnetic field $\pi$ is a ladder operator. In the Landau gauge $\vex A = (-B y, 0)$, $p_x$ is conserved and $\pi = p_x + B y+ip_y$.
Using periodic boundary conditions in the $x$ direction with length $L$, the spectrum of $h$ is given by eigenvectors $e^{-i 2\pi q x /L}\Phi(y-y_q)$, where $y_q = 2\pi q / BL$ is the guiding center,
\begin{equation}
\Phi = \Phi_{ns} = \left( \begin{array}{c} \phi_{n} \\ s \phi_{n-N} \end{array} \right),\quad n\geq N, s=\pm1,
\end{equation}
with energy $\epsilon_{ns}  = s\alpha (2B)^{N/2}\sqrt{(n-N+1)\cdots(n-1)n}$
and
\begin{equation}\label{eq:zero-rib}
\Phi = \Phi_{n0} = \left( \begin{array}{c} \phi_{n} \\ 0 \end{array} \right),\quad 0\leq n<N,
\end{equation}
with energy $\epsilon_{n0}=0$. Here $n$ and $q$ are integers and $\phi_n(y)$ is the eigenvector of the harmonic operator $(\pi-p_x)(\pi^\dagger-p_x)$ with eigenvalue $2nB$. The guiding center can range over the width $W$ of the sample, resulting in the degeneracy $0\leq q < D=WLB/(2\pi)$ of the Landau levels equal to the number of flux quanta in the system. For a uniform mass, the full energy spectrum of $H$ is then found as $\pm \sqrt{\epsilon_{ns}^2+|M|^2}$. For $N=1$ spectrum this reproduces the well-known result in graphene. 



\emph{Vortex in the ribbon geometry}.---%
Now we consider a vortex in $M$ of winding number $Q$. In the two dimensional plane the mass term is $M_1+iM_2=|M| e^{-iQ\theta}$ where $\theta$ denotes the azimuthal angle on the plane and $|M|$ vanishes at the vortex core. A topologically equivalent geometry is an annulus (or Corbino disk) where $|M|$ is constant in the sample and vanishes in the central hole.

\begin{figure}
\input{epsf}
\includegraphics[width=0.48\textwidth]{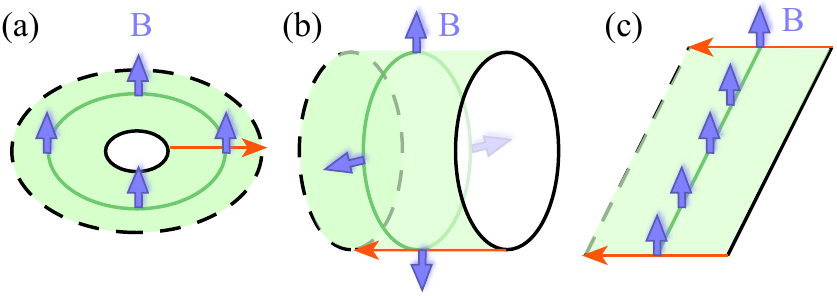}
\caption{(color online) Topologically equivalent geometries: Corbino disk (a), cylinder (b), and ribbon (c). In each geometry the bulk is shaded (green), the two edges are shown by thick solid and dashed (blue) lines, and the magnetic field is shown by large (blue) arrows. The `radial' direction connecting the two edges is shown by the thin (orange) arrow. Also a thick (green) line is shown to guide the eye on the orientation of the magnetic field in the bulk. In the thermodynamic limit, the edge shown by the dashed line is taken to infinity.\label{fig:geometry}}
\end{figure}

We may remove the position-dependent phase with a unitary transformation
$H \mapsto U^\dagger H U$, 
$
U = \left(\begin{array}{cccc}
    e^{-iQ\theta} & 0\\
    0 & \id
    \end{array}\right)
$,
at the expense of generating additional flux in the kinetic energy, that is 
\begin{equation}
  H \mapsto \left(\begin{array}{cccc}
    h({\vex A+\vex a}) & |M|   \\
    |M| & -h(\vex A)\end{array}\right),\label{eq:Ham'}
\end{equation}
where $\vex a=Q\gvex\nabla \theta$ is the vector potential corresponding to the additional flux tube of $Q$ quanta piercing the hole, $\oint\vex a\cdot d\vex r = 2\pi Q$. In the context of quantum Hall effect, the insertion of flux of $Q$ flux quanta was shown to transport $Q$ electrons between the two edges of the Corbino disk~\cite{Hal82a}. This adiabatic transport is the physical process leading to zero modes in our problem.

The spectrum of $H$ can be analyzed much more transparently in a ribbon geometry, which is topologically equivalent to the Corbino disk. It can in turn be flattened to a strip of length $L$ with the top and bottom edges identified, i.e. with the periodic boundary condition imposed along the $x$ direction~\cite{Lau81a}. Now   the vector potential $\vex a$ is simply a large gauge field, $\int_0^La_xdx=2\pi Q$, and we may choose the gauge $\vex a=(2\pi Q/L,0)$.


\emph{Selection rules}.---%
The entire effect of the vortex has now been reduced to a large gauge transformation in the Hamiltonian projected by $(1+\gamma_5)/2$, i.e. in the block $h(\vex A+\vex a)$. This results in the adiabatic transport of $Q$ electrons from one edge to the other, shifting the guiding center to $y_q=2\pi(q+Q)/BL$. Therefore, for these projected states, labeled as $|nqs+\rangle$, 
$
-Q\leq q < D-Q.
$
By contrast the states projected by $(1-\gamma_5)/2$ are labeled $|nqs-\rangle$ with $0\leq q <D$ as before.

The off-diagonal $M$ couples the projected states, but it satisfies the selection rule
\begin{eqnarray}
  \langle nqs+|M|n'q's'+\rangle &=&   \langle nqs-|M|n'q's'-\rangle = 0,\\
  \langle nqs+|M|n'q's'-\rangle &=& ss'\mu_{nn'}(Q) \delta_{qq'},\label{eq:S_rule}
\end{eqnarray}
with 
$
  \mu_{nn'}(Q)=\int\phi_{n}(y-y_Q)\phi_{n'}(y) d y
$.

\emph{Index}.---%
Assuming $Q>0$ without loss of generality, the full Hamiltonian in this basis is a direct sum of blocks with vanishing elements $\langle nqs+|M|n'qs-\rangle=0$ for $-Q\leq q <0$ and $D-Q\leq q <D$. Consequently, the set of states $|nqs-\rangle$ with $q\in[-Q,0)$ couple neither among themselves nor to any state $|n'q's'+\rangle$. The decoupling is also true for the states $|nqs+\rangle$ with $q\in[D-Q,D)$. Therefore, each of these state has the same energy as in the case when there is no vortex in $M$. However, the zero modes with $0\leq q <D-Q$ are now split away from zero energy due to the coupling among themselves and with non-zero energy states $n\geq N$. The zero modes $|nq0\pm\rangle$ are at the opposite edges of the ribbon. Thus, in the thermodynamic limit of a semi-infinite ribbon, the number of zero modes of the vortex Hamiltonian in the ribbon geometry in the presence of magnetic field is $NQ$.

The Hamiltonian~(\ref{eq:Ham}) anticommutes with the operator $\Gamma=\gamma_0\gamma_3$. The zero modes can thus be chosen to be eignestates of $\Gamma$ and $\mathrm{ind}\,H = \tr(\Gamma)$. The zero modes in the ribbon geometry satisfy $\Gamma|nq0\pm\rangle=\pm|nq0\pm\rangle$. Therefore, in the thermodynamic limit,
\begin{equation}\label{eq:index}
  \mathrm{ind}\,H=NQ.
\end{equation}
The index is a topological quantity that does not depend on smooth variations of geometry or choice of gauge. We conclude that the index is $NQ$ independent of the choices of geometry and gauge we made so far. This is our central result.

\emph{Vortex in the Corbino geometry}.---%
To make the connection to the original planar geometry clear, we also construct the spectrum of the Hamiltonian on the Corbino disk in the limit where the radius of the inner hole vanishes. The details of the derivations are given in the Supplemental Material~\cite{Note1}. In the symmetric gauge $\vex A =\frac12 Br\hat{\gvex\theta}$ with the polar coordinates $(r,\theta)$, we have the basis for the ladder operators $\pi \psi_{nm} = -i\sqrt{n}\psi_{n-1m+1}$ where $n\in\mathbb{N}\cup\{0\}$ is a radial quantum number and $m\geq -n$ is the angular momentum. The spectrum of $h(\vex A)$ is then found to be given with eigenvectors 
\begin{equation}
\Psi_{nms} = \left( \begin{array}{c}
i^N\psi_{nm} \\
s\psi_{n-Nm+N}
\end{array}\right),
\end{equation}
and energy $\epsilon_{ns}$ for $n\geq N$ and eigenvectors 
\begin{equation}
\Psi_{nm0} = \left( \begin{array}{c}
\psi_{nm} \\ 
0
\end{array}\right),
\end{equation}
with energy $\epsilon_{n0}=0$ for $0\leq n <N$. One can show that for the projection $h|_{m}$ onto the subspace with fixed $m$
\begin{equation}
\dim\ker h|_{m} = N \Theta(m) + (N+m) \Theta(m+N) \Theta(-m)
\end{equation}
with $\Theta$ the step function.

When the mass is nonzero and carries a vortex of winding $Q$, the basis states $|nms+\rangle = (\Psi_{mns},0)^\intercal$ and $|nms-\rangle = (0,\Psi_{mns})^\intercal$ get coupled with the following selection rules
\begin{eqnarray}
  \langle nms+|M|n'm's'+\rangle &=&   \langle nms-|M|n'm's'-\rangle = 0,\\
  \langle nms+|M|n'm's'-\rangle &=& \tilde\mu_{nn'm}(Q,ss') \delta_{mm'-Q},
\end{eqnarray}
where the overlap $\tilde\mu$ depends only on the sign $ss'$. Using these selection rules and the fact that $\Gamma|mn0\pm\rangle = \pm |nm0\pm\rangle$, one can show that the zero-energy states can only split if they are coupled with opposite projections (see Supplemental Material~\cite{Note1}). For example, for $N=3, Q=2$, the number of zero modes for $m=\cdots,-3, -2, -1, 0, 1, 2, \dots$ are, respectively, $\cdots,0,1,2,2,1,0,\dots$ adding up to $NQ=6$. In general
\begin{equation}\label{eq:N0}
\mathrm{ind}\,H = \sum_m \left( \dim\ker h|_m - \dim\ker h|_{m-Q} \right) = NQ,
\end{equation}
as before. 

\emph{General field profile}.---%
Our analysis so far has been for a constant magnetic field. However, the arguments leading to Eqs.~(\ref{eq:index}) and~(\ref{eq:N0}) suggest that our result may be more generally applicable. Indeed, the zero-energy states of $h(\vex A)$ survive for a general profile of the magnetic field. For $N=1$ their number equals  the total (integer) flux enclosed by the system~\cite{AhaCas79a}. For $N>1$, this result can be generalized by writing $A_x-iA_y = -2i\partial_z\varphi$ in the Coulomb gauge, where $z=x+iy$ and $\varphi$ is the solution to the Laplace equation $\nabla^2\varphi = B$. So, $\pi=2(\partial_{\bar z} +\partial_{\bar z}\varphi)$.
The zero modes are then found to be 
\begin{equation}
\left(\begin{array}{c}
e^{-\varphi(z,\bar z)}p(\bar{z})f(z) \\
0
\end{array}
\right),
\end{equation}
where $p$ is a polynomial of order $0\leq n<N$ and $f$ is an entire function of $z$, $\partial f/\partial{\bar z} = 0$. A basis can be chosen as $p(\bar z) = \bar{z}^n$, $f(z) = z^{m+n}$. Then, regularity at the origin yields $m\geq -n$. Here $m$ is the angular momentum and $n$ is a radial quantum number, as before. This basis is not orthogonal; however, it has the same order as the one chosen for the uniform field in the Corbino geometry. 

For the general field, the existence of the gap cannot be assured and needs to be explicitly checked for a given profile. If the gap does not close, $\dim\ker h|_{m}$ is still well defined and the same as that in constant field. The selection rules remain the same as well.  Therefore the number of zero modes is still given by Eq.~(\ref{eq:N0}).


\emph{Discussion}.---%
The zero modes $\Psi_{i0}(\vex r)$, $i=1,\cdots,N|Q|$, carry a valley spin texture, $\Psi_{i0}^\dagger(\vex r)\gvex\Sigma\Psi_{i0}(\vex r)$, where $\gvex\Sigma=(\gamma_0,i\gamma_0\gamma_5,\gamma_5)$. Since the Hamiltonian anticommutes with $\Gamma=\gamma_0\gamma_3$ and $[\gamma_5,\Gamma]=0$, the ground-state value $\left<\Sigma_z\right>$ is entirely determined by the zero modes:
$$
\left<\Sigma_z\right> = -\frac12\sum_i\left<\Psi_{i0}|\gamma_5|\Psi_{i0}\right>.
$$
The magnitude and the direction of $\left<\Sigma_z\right>$ is determined by $B$. In particular $\sgn \left<\Sigma_z\right> \propto \sgn B$. Therefore as the vortex moves, $\left<\Sigma_z\right>$ changes with $B$ and a valley current flux $\gvex\nabla\cdot j^v\propto \dot{\vex r}\cdot \gvex\nabla B$ is produced. Assuming a symmetric distribution of current, this amounts to a valley current normal to the motion of the vortex. The magnitude of the valley current depends on the number of zero modes. 

The detailed form of the wavefunctions depends, of course, on the choice of geometry. However, as we showed explicitly, as long as the topology is preserved, the number of zero modes is determined by an index theorem connecting the analytical and topological indices of the generalized Dirac operator. Our result agrees with previous studies for $N=1$~\cite{JacRos81a} and for $N=2$~\cite{LuHer12a} without a magnetic field. Our extension to the case of finite magnetic fields and general $N$ shows that the index is independent of the external magnetic field as long as no gaps are closed. It also determines the assignment of quantum numbers (charge, spin) to vortices and skyrmions~\cite{LuHer12a,HerLuRoy12}, and can have direct signatures in transport and critical behavior of systems where generalized Dirac Hamiltonians govern the dynamics of low-energy excitations, notably multi-layer graphene in magnetic field.

\acknowledgements
The authors acknowledge useful discussions with Fernando de Juan, Herb Fertig, Igor Herbut, Arijit Kundu, Ganpathy Murthy, and Bitan Roy. This work was supported by the NSF through Grant No. DMR-1005035 and by the College of Arts and Sciences, Indiana University--Bloomington.

\bibliographystyle{physre}

\begin{thebibliography}{10}

\bibitem{JacReb76a}
R.~Jackiw and C.~Rebbi,
\newblock \prd\ {\bf 13}, 3398 (1976).

\bibitem{SuSchHee79a}
W.~P. Su, J.~R. Schrieffer, and A.~J. Heeger,
\newblock \prl\ {\bf 42}, 1698 (1979).

\bibitem{JacSch81a}
R.~Jackiw and J.~R. Schrieffer,
\newblock Nuclear Physics B {\bf 190}, 253 (1981).

\bibitem{NieSem83a}
A.~J. Niemi and G.~W. Semenoff,
\newblock \prl\ {\bf 51}, 2077 (1983).

\bibitem{SemSemZho08a}
G.~W. Semenoff, V.~Semenoff, and F.~Zhou,
\newblock \prl\ {\bf 101}, 087204 (2008).

\bibitem{LuHer12a}
C.-K. Lu and I.~F. Herbut,
\newblock \prl\ {\bf 108}, 266402 (2012).

\bibitem{Note1}
See Supplemental Material for definition and detailed calculation of the index, several examples, and an exact solution in the vortex-core limit.

\bibitem{AtiSin63a}
M.~F. Atiyah and I.~M. Singer,
\newblock Bull. Amer. Math. Soc. {\bf 69}, 422 (1963).

\bibitem{HouChaMud07a}
C.-Y. Hou, C.~Chamon, and C.~Mudry,
\newblock \prl\ {\bf 98}, 186809 (2007).

\bibitem{WeeFra10a}
C.~Weeks and M.~Franz,
\newblock \prb\ {\bf 81}, 085105 (2010).

\bibitem{Her06b}
I.~F. Herbut,
\newblock \prl\ {\bf 97}, 146401 (2006).

\bibitem{Kha12a}
M.~Kharitonov,
\newblock \prl\ {\bf 109}, 046803 (2012).

\bibitem{ZhaMinMac12a}
F.~Zhang, H.~Min, and A.~H. MacDonald,
\newblock \prb\ {\bf 86}, 155128 (2012).

\bibitem{Her10b}
I.~Herbut,
\newblock \prl\ {\bf 104}, 066404 (2010).

\bibitem{GhaWil12a}
P.~Ghaemi and F.~Wilczek,
\newblock Physica Scripta {\bf 146}, 014019 (2012).

\bibitem{HosZar12a}
M.~V. Hosseini and M.~Zareyan,
\newblock \prl\ {\bf 108}, 147001 (2012).

\bibitem{MurVaf13a}
J.~M. Murray and O.~Vafek,
\newblock arXiv:1312.6831  (2013).

\bibitem{LozSok08a}
Y.~Lozovik and A.~Sokolik,
\newblock JETP Letters {\bf 87}, 55 (2008).

\bibitem{MinBisSu08b}
H.~Min, R.~Bistritzer, J.-J. Su, and A.~H. MacDonald,
\newblock \prb\ {\bf 78}, 121401 (2008).

\bibitem{ZhaJog08b}
C.-H. Zhang and Y.~N. Joglekar,
\newblock \prb\ {\bf 77}, 233405 (2008).

\bibitem{SerWebFra08a}
B.~Seradjeh, H.~Weber, and M.~Franz,
\newblock \prl\ {\bf 101}, 246404 (2008).

\bibitem{Ser12b}
B.~Seradjeh,
\newblock \prb\ {\bf 86}, 121101(R) (2012).

\bibitem{PerNeiHam13a}
A.~Perali, D.~Neilson, and A.~R. Hamilton,
\newblock \prl\ {\bf 110}, 146803 (2013).

\bibitem{JacRos81a}
R.~Jackiw and P.~Rossi,
\newblock Nuclear Physics B {\bf 190}, 681 (1981).

\bibitem{Her07a}
I.~F. Herbut,
\newblock \prl\ {\bf 99}, 206404 (2007).

\bibitem{Ser08a}
B.~Seradjeh,
\newblock Nuclear Physics B {\bf 805}, 182 (2008).

\bibitem{SerWeeFra08a}
B.~Seradjeh, C.~Weeks, and M.~Franz,
\newblock \prb\ {\bf 77}, 033104 (2008).

\bibitem{SerMooFra09a}
B.~Seradjeh, J.~E. Moore, and M.~Franz,
\newblock \prl\ {\bf 103}, 066402 (2009).

\bibitem{Wei81a}
E.~J. Weinberg,
\newblock \prd\ {\bf 24}, 2669 (1981).

\bibitem{Her10a}
I.~F. Herbut,
\newblock \prb\ {\bf 81}, 205429 (2010).

\bibitem{HouChaMud10a}
C.-Y. Hou, C.~Chamon, and C.~Mudry,
\newblock \prb\ {\bf 81}, 075427 (2010).

\bibitem{CheLiOng09a}
J.~G. Checkelsky, L.~Li, and N.~P. Ong,
\newblock \prl\ {\bf 100}, 206801 (2008); 
\newblock \prb\ {\bf 79}, 115434 (2009).

\bibitem{AhaCas79a}
Y.~Aharonov and A.~Casher,
\newblock \pra\ {\bf 19}, 2461 (1979).

\bibitem{McCFal06a}
E.~McCann and V.~I. Fal'ko,
\newblock \prl\ {\bf 96}, 086805 (2006).

\bibitem{Hal82a}
B.~Halperin,
\newblock \prb\ {\bf 25}, 2185 (1982).

\bibitem{Lau81a}
R.~B. Laughlin,
\newblock \prb\ {\bf 23}, 5632 (1981).

\bibitem{HerLuRoy12}
I.~F. Herbut, C.-K. Lu, and B. Roy,
\newblock \prb\ {\bf 86}, 075101 (2012).

\end{thebibliography}

\begin{thebibliography}{10}

\bibitem{peeters} F. Geerinckx, F. M. Peeters, and J. T. Devreese, J. App. Phys. {\bf 68}, 3435 (1990).

\bibitem{egger} A. De Martino and R. Egger, Semicond. Sci. Technol. {\bf 25}, 034006 (2010). 

\bibitem{LuHerbutPRL} C.-K. Lu and I.F. Herbut, Phys. Rev. Lett. {\bf 108}, 266402 (2012).

\bibitem{Babak} B. Seradjeh, Nuclear Phys. B {\bf 805}, 182 (2008)

\bibitem{IgorLu} I. F. Herbut and C.-K. Lu, Phys. Rev. B {\bf 83}, 125421 (2011).

\bibitem{Lu_JphysA} C.-K. Lu and I. F. Herbut, J. Phys. A {\bf44}, 295003 (2011).

\end{thebibliography}

\renewcommand{\thefigure}{S\arabic{figure}}
\setcounter{figure}{0}
\renewcommand{\theequation}{S\arabic{equation}}
\setcounter{equation}{0}

\section{Supplemental Material}

\subsection{Chiral symmetry and index}
The generalized Dirac-vortex Hamiltonian in the coordinate basis reads
\begin{equation}
  H=\left(\begin{array}{cc}
  h({\bf A}) & Me^{-iQ\theta} \\
  Me^{iQ\theta} & -h({\bf A})
  \end{array}\right).\label{H_vortex}
\end{equation} 
The diagonal parts correspond to the kinetic energy terms for $K_+$ and $K_-$ valleys in the representation.
Since the general vortex Hamiltonian anticommutes with the chiral symmetry operator $\Gamma=\gamma_0\gamma_3=\left(\begin{array}{cc} \sigma_3 & 0 \\ 0 & -\sigma_3 \end{array} \right)$ for any $N$ and $Q$, we may define the index associated with it,
\begin{equation}
	\mathrm{ind}\,H\equiv{\rm Tr}(\Gamma) = \nu_+ - \nu_-,
\end{equation}
where $\nu_\pm$ are the number of zero modes of $H$ that are eigenstates of $\Gamma$ with eigenvalue $\pm1$.

To see this, we first unitarily map the original basis $(u_+,v_+,v_+,u_-)\mapsto(u_+,u_-,v_+,v_-)$. In the new basis, the Hamiltonian takes the form
\begin{equation}
H = \left(\begin{array}{cc} 0 & D^\dagger \\ D & 0 \end{array}\right),
\end{equation}
with
\begin{equation}
D = \left( \begin{array}{cc} \alpha\pi^N & Me^{-iQ\theta} \\ M e^{iQ\theta} & - \alpha{\pi^\dagger}^N \end{array} \right).
\end{equation}
In this basis, $\Gamma = \sigma_3\otimes\id$. The index of $H$ is then defined in terms of the analytical index of $D$,
\begin{equation}
\mathrm{ind}\, H = {\rm dim\, ker}\, D - {\rm dim\, ker}\, D^\dagger.
\end{equation}
Since 
$$
H^2 = \left( \begin{array}{cc} D^\dagger D & 0 \\ 0 & DD^\dagger \end{array} \right),
$$
there is a one-to-one mapping between the zero modes $\psi_0$ and $\phi_0$, respectively, of $D$ and $D^\dagger$ to the zero modes $\Psi_0 = (\psi_0,0)^\intercal$ and $\Phi_0 = (0,\phi_0)^\intercal$ of $H$, which are, respectively, eigenstates of $\Gamma$ with eigenvalues $1$ and $-1$. Therefore, ${\rm dim\, ker}\, D = \nu_+$ and ${\rm dim\, ker}\, D^\dagger = \nu_-$. This completes the proof.

\subsection{Landau levels in the symmetric gauge}

The  momentum operator 
$$
\pi=(p_x-A_x)-i(p_y-A_y),
$$
with symmetric gauge $(A_x,A_y)=\frac{1}{2}B(-y,x)$. We assume $B>0$ without loss of generality. Writing $\rho\equiv \frac12Br^2$, we have
$$
\pi=-ie^{i\theta}\sqrt{\rho}\left(\frac{\partial}{\partial \rho}+\frac{i}{2\rho}\frac{\partial}{\partial\theta}+\frac{1}{2}\right).
$$
It acts as the ladder operator to the basis functions, 
\begin{equation}
  \pi\psi_{nm} = -i\sqrt{n}\psi_{n-1m+1}\:,\label{basis}
\end{equation} 
in which the basis functions are in the form of confluent hypergeometric functions,\cite{peeters,egger}
\begin{equation}
  \psi_{nm} = C_{nm}\ e^{im\theta}e^{-\rho/2}\rho^{m/2}\ {_1F_1}(-n,m+1,\rho)\:,\label{hyper1}
\end{equation}
with the normalization constant
\begin{equation}
  C_{nm}=\frac{1}{\ell}\left[\frac{1}{m!}
  \left(\begin{array}{cccc}
  n+m\\
  n
  \end{array}\right)\right]^{1/2}\:,
\end{equation} 
for $m\geq 0$, and
\begin{equation}
  \psi_{nm} = C_{nm}\ e^{im\theta}e^{-\rho/2}\rho^{-m/2}\ {_1F_1}(-n-m,1-m,\rho)\:,\label{hyper2}
\end{equation} 
with
\begin{equation}
  C_{nm}=\frac{1}{\ell}\left[\frac{1}{|m|!}
  \left(\begin{array}{cccc}
  n\\
  |m|
  \end{array}\right)\right]^{1/2}\:,
\end{equation} 
for $m<0$. In fact, Eq.~(\ref{basis}) can be verified using the following properties of hypergeometric functions,
\begin{align}
   \frac{d}{d\rho}{_1F_1}(a,b,\rho) &= \frac{a}{b}{_1F_1}(a+1,b+1,\rho)\:, \\
   [\rho\frac{d}{d\rho}+(b-1)]&{{}_1F_1}(a,b,\rho) = (b-1)\ {_1F_1}(a,b-1,\rho)\:.
\end{align}

From the expressions in Eqs.~(\ref{hyper1}) and~(\ref{hyper2}), and the fact that the confluent 
hypergeometric functions $_1F_1$ reduce to polynomial if the first argument is a nonpositive integer, the allowed angular momenta for a given integer $n\geq 0$ are $m\geq -n$. Now one can readily show that the eigenstates of the Dirac-Landau Hamiltonian $h({\bf A})= \alpha \left( \begin{array}{cc} 0 & {\pi^{\dag}}^N \\ {\pi}^N & 0 \end{array}\right)$ are,
\begin{equation}
  \Psi_{nms}=\frac{1}{\sqrt{2}}
  \left(\begin{array}{c}
  i^N\psi_{nm}\\
  s\ \psi_{n-Nm+N}
  \end{array}\right)\:,\label{basis1}
\end{equation} 
with eigenvalues $\epsilon_{ns}=s\alpha(2B)^{N/2}\sqrt{\frac{n!}{(n-N)!}}$. Here, $s=\pm1$ and $n\geq N$. The the zero-energy states are
\begin{equation}
  \Psi_{nm0}=
  \left(\begin{array}{c}
  \psi_{nm}\\
  0
  \end{array}\right),\label{basis2}
\end{equation} 
for $0\leq n<N$. Note that $\sigma_3\Psi_{nm\pm}=\Psi_{nm\mp}$ and $\sigma_3\Psi_{nm0}=+\Psi_{nm0}$.

\subsection{Zero modes and the index}

We may project the Hamiltonian onto the Landau level (LL) basis, obtaining an infinite-dimension matrix presentation of the Hamiltonian. Working with the basis functions that are eigenstates of angular momentum, it can be seen that in the presence of a mass vortex the subspaces
\begin{eqnarray}
\mathcal{H}_{mQ} &=& \mathcal{H}_{m}^+ \oplus \mathcal{H}_{m+Q}^-,~\mathrm{where}\\
\mathcal{H}_{m}^+ &=& \mathrm{span}\left\{
	\left(\begin{array}{c}
  	\Psi_{nm0}\\
	0
  	\end{array}\right),
	\left(\begin{array}{c}
  	\Psi_{nms}\\
  	0
  	\end{array}\right)
	\right\},
	\\
\mathcal{H}_{m+Q}^- &=& \mathrm{span}\left\{
	\left(\begin{array}{c}
  	0\\
  	\Psi_{nm+Q0}
  	\end{array}\right),
	\left(\begin{array}{c}
  	0\\
  	\Psi_{nm+Qs}
  	\end{array}\right)\right\},
\end{eqnarray}  
are decoupled for different values of $m\in\mathbb{Z}$. (Recall $n\geq -m$.) Thus the matrix representation of $H$ is {\it reducible}. We denote the basis states in $\mathcal{H}_m^\pm$ with $|nm0\pm\rangle$ and $|nms\pm\rangle$.

Thus
$$
{\rm ind}\,H = \sum_m{\rm Tr}_{\mathcal{H}_{mQ}}(\Gamma),
$$
where the sum is over partial  traces in the subspace $\mathcal{H}_{mQ}$. Since $\Gamma|nm0\pm\rangle = \pm|nm0\pm\rangle$ and $\Gamma|nms\pm\rangle = |nm-s\pm\rangle$, the partial trace is found to be
\begin{equation}
	{\rm Tr}_{\mathcal{H}_{mQ}}(\Gamma)={\rm dim}\:{\rm ker}(h|_{m+Q})-{\rm dim}\:{\rm ker}(h|_{m}),\label{general_ind}
\end{equation} 
where $h|_{m}$ is the projection of the Hamiltonian $h({\bf A})$ onto the subspace with a given angular momentum $m$.

It is straightforward to see that 
\begin{equation}
{\rm dim}\ {\rm ker} (h|_m)= N\Theta(m)+(N+m)\Theta(m+N)\Theta(-m).
\end{equation}
Taking $Q>0$ without loss of generality, two situations arise. First, $Q>N$. Then
\begin{equation}
{\rm Tr}_{\mathcal{H}_{mQ}}(\Gamma) = \begin{cases} 
	N+Q+m 	& -N-Q\leq m\leq -Q, \\  
	N 		& -Q\leq m \leq-N, \\ 
	-m 		& -N \leq m\leq 0, \\ 
	0 		& {\rm otherwise}. \end{cases}
\end{equation}
Thus,
\begin{equation}
	{\rm ind}\,H={N(N-1)}+N(Q-N+1)=NQ. 
\end{equation}
Second, $Q<N$. Then
\begin{equation}
{\rm Tr}_{\mathcal{H}_{mQ}}(\Gamma) = \begin{cases} 
	N+Q+m 	& -N-Q\leq m\leq -N, \\  
	Q 		& -N\leq m \leq-Q, \\ 
	-m 		& -Q \leq m\leq 0, \\ 
	0 		& {\rm otherwise}. \end{cases}
\end{equation}
Again,
\begin{equation}
	{\rm ind}\,H = Q(Q-1)+Q(N-Q+1) = NQ.
\end{equation}

In the followings, we shall demonstrate with a few examples that the number of zero-energy eigenvalues of $H$ is index$(H)$.

\subsubsection{Case of $N=Q=1$}

\begin{figure}
\input{epsf}
\includegraphics[width=0.48\textwidth]{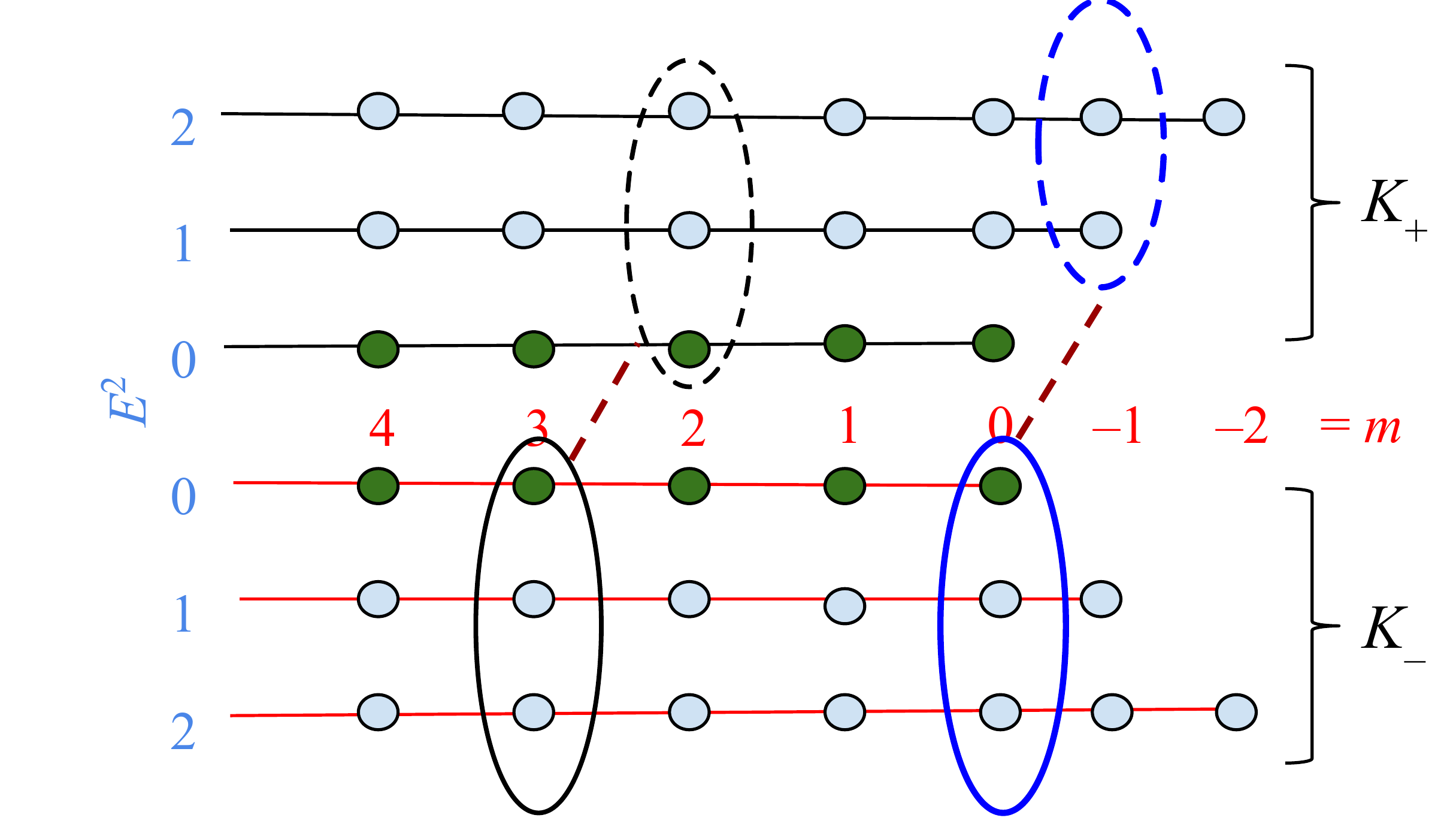}
\caption{Landau level basis for $N=1$. In each column the states in the upper (lower) half are eigenstates of $h(\bf A)$ associated with $K_+$ ($K_-$) valley. Dimensionless energy $E^2=n$ and angular momentum $m$ are used to label these states. Green (blue) circles indicate that there is (are) one (two) state(s) associated with the given $n$ corresponding to $\pm E$. Brown dashed lines denote the coupling between the Landau levels of different valleys as a result of the vortex with $Q=1$. One zero mode is found from the subspace included in the blue ovals while the black one, like the rest of the sectors, does not yield a zero mode.}\label{single}
\end{figure}

The LL's in the subspace $\mathcal H_{-1,1}$ are marked by the blue ovals in Fig.\ \ref{single}. We shall demonstrate that the submatrix $H_1$ for this subspace does yield one zero eigenvalue. Now for simplicity we pick up a subset of 5 LL's from this sector that lie most close to zero energy,
\begin{equation}
  \{\left(\begin{array}{cccc}
  0\\
  0\\
  \Psi_{0,0,0}
  \end{array}\right)\:,\ 
  \left(\begin{array}{cccc}
  0\\
  0\\
  \Psi_{1,0,\pm}
  \end{array}\right)\:,\ 
  \left(\begin{array}{cccc}
  \Psi_{1,-1,\pm}\\
  0\\
  0
  \end{array}\right)\}_{N=1}\:,
\end{equation}
where the subscript is to remind that the LL's here correspond to $N=1$ in Eq.\ (\ref{basis1}) and (\ref{basis2}). Then the submatrix has the following form,
\begin{equation}
  H_1=\left(\begin{array}{ccccc}
  0 & 0 & 0 & \alpha_1 & \alpha_1\\
  0 & \Omega_1 & 0 & \beta_1 & \gamma_1\\
  0 & 0 & -\Omega_1 & \gamma_1 & \beta_1\\
  \alpha_1^* & \beta_1^* & \gamma_1^* & \Omega_1 & 0 \\
  \alpha_1^* & \gamma_1^* & \beta_1^* & 0 & -\Omega_1
  \end{array}\right)\:,\label{sub_h1}
\end{equation}
in which the nonzero couplings represented by the Greek letters are only between states of opposite valleys. It should be noted that in the first row the matrix elements 
\begin{equation}
  \alpha_1=\int  rdr d\theta\ [\Psi_{1,-1,\pm}]^{\dag}M(r)e^{-i\theta}\Psi_{0,0}\:,
\end{equation}
representing the coupling between zero-energy LL of K$_+$  and first LL's in K$_-$ are identical. This is due to the chiral symmetry of $H$. $\beta_1$ and $\gamma_1$ represent the two different couplings between the first LL's in both valleys. $\Omega_1$ refers to the cyclotron energy for single layer graphene.

It can be shown that the ansatz column $(w,y,y,z,-z)^T$ is the eigenstate of zero eigenvalue of the approximating matrix $H_1$. The unknowns can be easily calculated, giving $y/z=(\gamma_1-\beta_1)/\Omega_1$ and $w/z=-\Omega_1/\alpha_1^*+(\beta_1^*+\gamma_1^*)(\beta_1-\gamma_1)/\Omega_1$. Now the approximating wavefunction for the zero mode reads, 
\begin{equation}
  |\psi^0\rangle=\left(\begin{array}{c}
   0\\	  
  -z\psi_{0,0}\\
  \frac{w}{\sqrt{2}}\psi_{0,0}+iy\psi_{1,0}\\
  0
  \end{array}\right)\:,
\end{equation}
where the same (opposite) signs appeared in the pair of $y$'s ($z$'s) in the components are crucial such that the above zero mode is also eigenstate of chirality operator $\Gamma$. As more of the rest states are considered, the dimension of matrix increases by an even number as pairs of states with same LL index $n$ are included. We conclude that the number of zero eigenvalue of the entire matrix is indeed given by the sum,  
\begin{equation}
  \sum_{m=-\infty}^{\infty}{\rm Tr}_{\mathcal H_{m1}}(\Gamma)=1,
\end{equation}
in which only the partial trace for $m=-1$ is nonzero as shown by $H_1$.

\subsubsection{Case of $N=2$ and $Q=1$}
\begin{figure}
\input{epsf}
\includegraphics[width=0.48\textwidth]{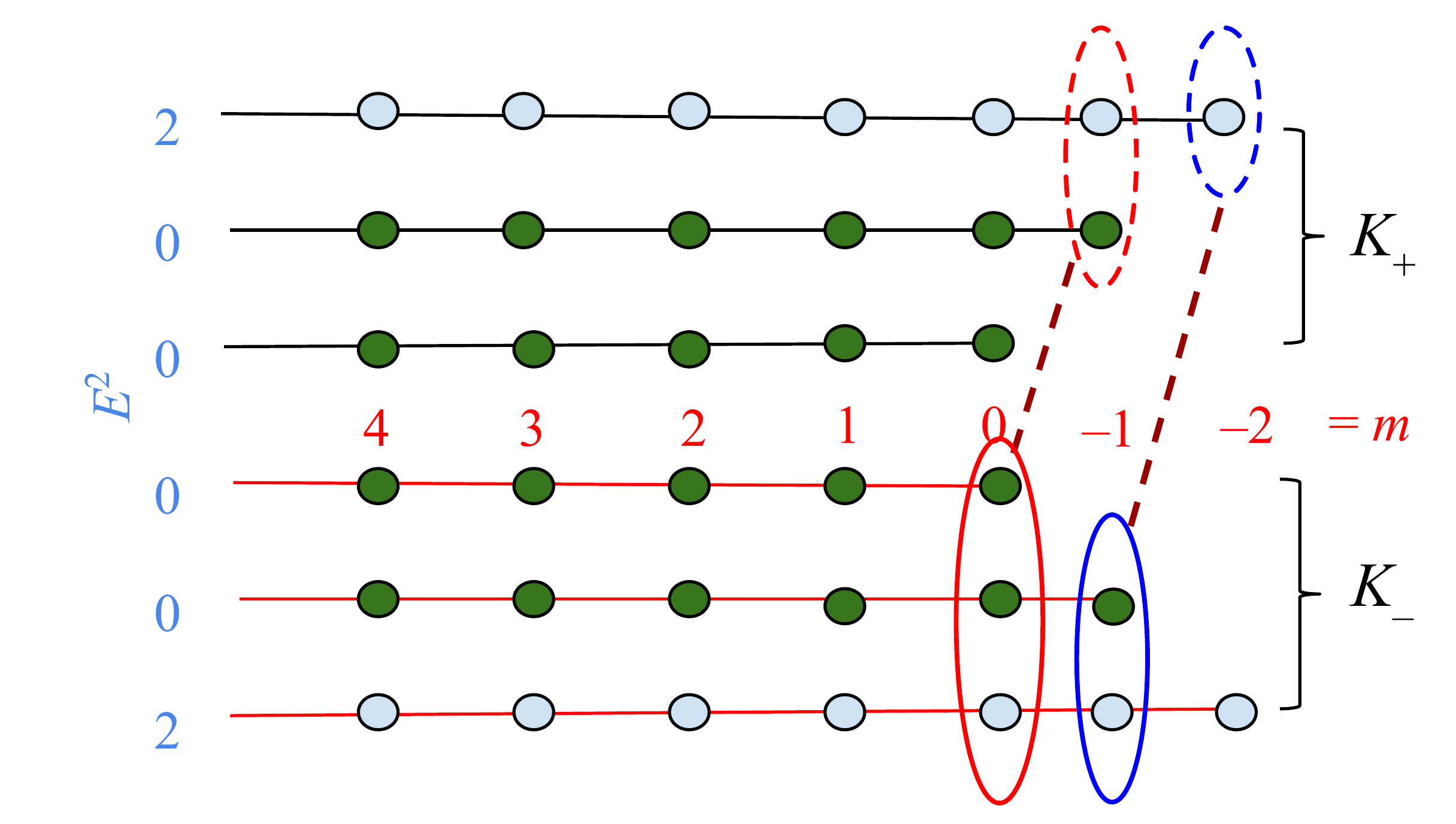}
\caption{Landau level basis for $N=2$ corresponding to the Bernal-stacked bilayer graphene. Labels have the same meaning as in Fig.~\ref{single}. The blue ovals here are similar to its counterparts in Fig.~\ref{single}, and one zero mode is found. The red ones contain three zero modes without the vortex, one of which is unpaired when there is a vortex with $Q=1$.}\label{bilayer}
\end{figure}

The same argument can be repeated in the case with $N=2$ and $Q=1$. As shown in Fig.\ \ref{bilayer}, the fact that $\pi^2|n=1,m\geq -1\rangle=0$ result in the doubling of zero-energy LL subspace, which means that ${\rm dim}\ {\rm ker}(h|_m)$ is {\it two} for $m\geq 0$, {\it one} for $m=-1$, and {\it zero} otherwise. We first focus on the sector $\mathcal H_{-2,1}$ indicated by the blue ovals, which is similar to the sector indicated by the blue ovals in Fig.\ \ref{single}. It can be shown that the corresponding submatrix has identical structure as $H_1$ in Eq.\ (\ref{sub_h1}). Then one can see that the zero mode solution is of the form,
\begin{equation}
  |\psi^0_1\rangle=
  \sum_{n\geq 1}c_n\left(\begin{array}{c}
  0\\
  0\\
  \psi_{n,-1}\\
  0
  \end{array}\right)
  +\sum_{n\geq 0}d_n
  \left(\begin{array}{c}
  0\\
  \psi_{n,0}\\
  0\\
  0\\
  \end{array}\right)\:,\label{first0}
\end{equation}
where the coefficients $c_n$ and $d_n$ depend on the cyclotron energy in bilayer and the overlap integrals between LL's.

The second zero mode will be shown to come from the subspace $\mathcal H_{-1,1}$ indicated by the red ovals which has ${\rm dim}\ {\rm ker}(h|_{-1})=1$ (red dashed oval) and ${\rm dim}\ {\rm ker}(h|_0)=2$ (red solid oval). We start with considering the five states around zero energy,
\begin{equation}
  \{\left(\begin{array}{cccc}
  0\\
  0\\
  \Psi_{0,0,0}
  \end{array}\right)\:,
  \left(\begin{array}{cccc}
  0\\
  0\\
  \Psi_{1,0,0}
  \end{array}\right)\:, 
  \left(\begin{array}{cccc}
  \Psi_{1,-1,0}\\
  0\\
  0
  \end{array}\right)\:,
  \left(\begin{array}{cccc}
  \Psi_{2,-1,\pm}\\
  0\\
  0
  \end{array}\right)\}_{N=2}\:.
\end{equation}
The submatrix with nonzero matrix elements denoted by Greek letters now reads,
\begin{equation}
  H_2=\left(\begin{array}{ccccc}
  0 & 0 & \gamma_2 & \alpha_2 & \alpha_2\\
  0 & 0 & \delta_2 & \beta_2 & \beta_2\\
  \gamma_2^* & \delta_2^* & 0 & 0 & 0\\
  \alpha_2^* & \beta_2^* & 0 & 2\Omega_2 & 0 \\
  \alpha_2^* & \beta_2^* & 0 & 0 & -2\Omega_2
  \end{array}\right)\:,\label{sub_h2}
\end{equation}
where $\Omega_2$ denotes the cyclotron energy in bilayer graphene. We find that the ansatz column $(\delta_2^*w,-\gamma_2^*w,0,z,-z)^T$ correspond to the zero eigenvalue of the approximating submatrix $H_2$, and it is easy to show that $z/w=\frac{\beta_2^*\gamma_2^*-\alpha_2^*\delta_2^*}{2\Omega_2}$. Consequently, the second zero mode is of the form,
\begin{equation}
  |\psi^0_2\rangle=\sum_{n\geq 0} c'_n
  \left(\begin{array}{c}
  0\\
  0\\
  \psi_{n,0}\\
  0
  \end{array}\right)+\sum_{n\geq 0}d'_n
  \left(\begin{array}{c}
  0\\
  \psi_{n,1}\\
  0\\
  0\\
  \end{array}\right)\:.\label{second0}
\end{equation}
In the absence of field, the pair of zero modes in Eq.\ (\ref{first0}) and (\ref{second0}) are related by an antilinear operator which guarantees the doubling of zero mode.\cite{LuHerbutPRL}

\subsubsection{Case of $N=Q=2$}
\begin{figure}
\input{epsf}
\includegraphics[width=0.48\textwidth]{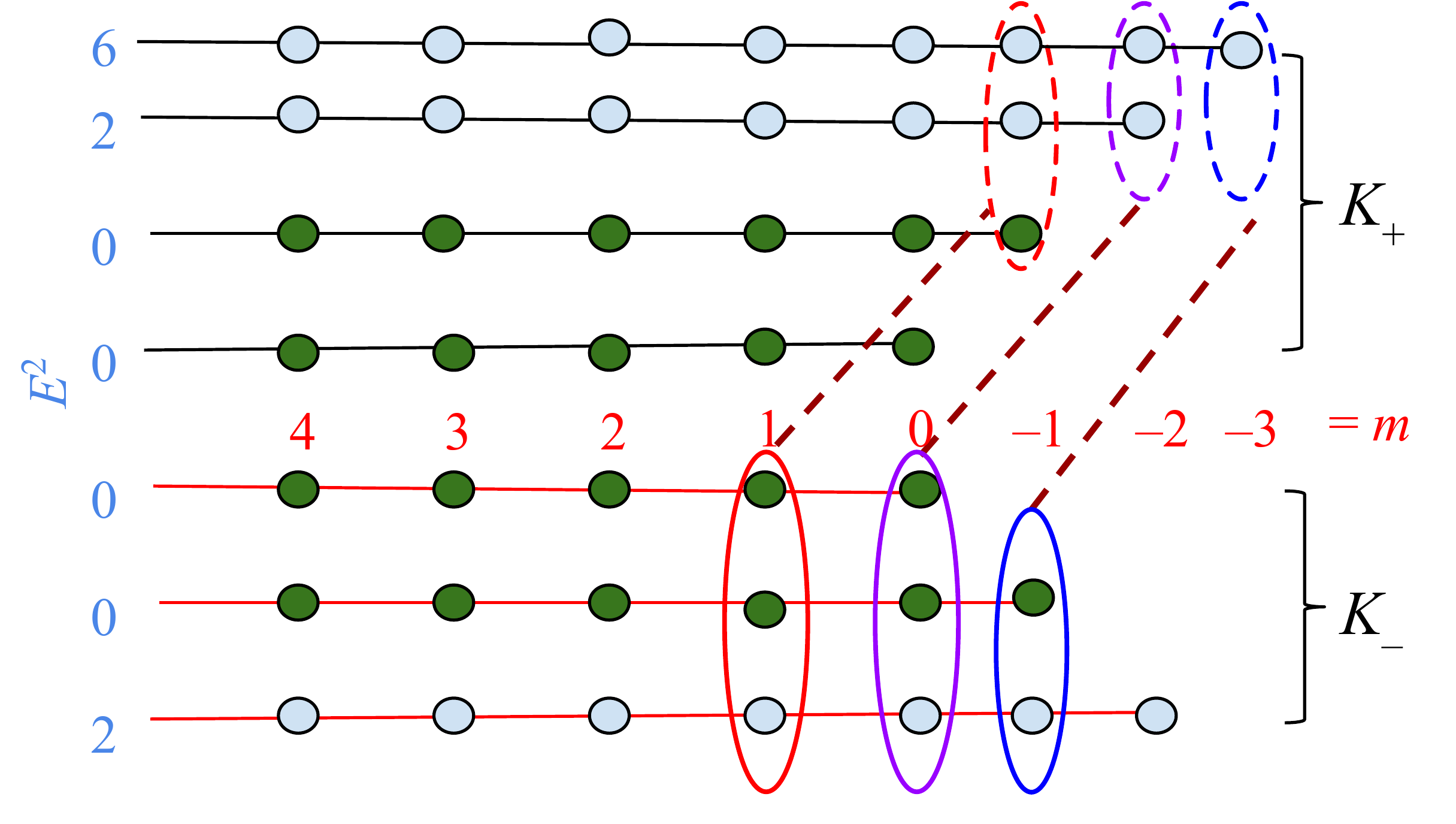}
\caption{Same as Fig.~\ref{bilayer} but the vortex has winding number $Q=2$. The blue and red ovals share the same structures with their counterparts in Fig.~\ref{bilayer}, and two zero modes are found from them. The purple ovals, on the other hand, contain two zero modes without the vortex, both of which remain unpaired in the presence of the vortex. Thus, a total of four zero modes are found here.}\label{bilayer_q2}
\end{figure}

Here the case for a double vortex in double layer is worthwhile mentioning. The couplings of LL's due to the double vortex is shown in Fig.\ \ref{bilayer_q2}. The subspace $\mathcal H_{-3,2}$ indicated by the blue and $\mathcal H_{-1,2}$ indicated by the red ovals can be shown to relate to the submatrices $H_1$ in Eq.\ (\ref{sub_h1}) and $H_2$ in Eq.\ (\ref{sub_h2}), respectively. So each of them results in one zero mode. The subspace $\mathcal H_{-2,2}$ indicated by the purple ovals is different in that ${\rm dim}\ {\rm ker}(h|_{-2})=0$ (purple dashed oval) and ${\rm dim}\ {\rm ker}(h|_0)=2$ (purple solid oval), and we shall show that the number of zero modes is indeed two. The submatrix associated with this sector reads,
\begin{equation}
  H_3=\left(\begin{array}{cccccc}
  0 & 0 & 0 & 0 & \alpha_3 & \alpha_3\\
  0 & 0 & 0 & 0 & \beta_3 & \beta_3\\
  0 & 0 & 2\Omega_2 & 0 &\gamma_3 & \delta_3\\
  0 & 0 & 0 & -2\Omega_2 & \delta_3 &\gamma_3\\
  \alpha_3^* & \beta_3^* & \gamma_3^*& \delta_3^* & 2\Omega_2 &0\\
  \alpha_3^* & \beta_3^* & \delta_3^*& \gamma_3^* & 0 &-2\Omega_2
  \end{array}\right)\:.
\end{equation}
One can show that the two ansatz columns, $(w_1,0,y_1,y_1,z_1,-z_1)^T$ and $(0,w_2,y_2,y_2,z_2,-z_2)^T$ yield zero eigenvalue. Those unknowns are also determined by the overlap integrals and $\Omega_2$. Consequently, the present case of $N=Q=2$ yields four zero modes.

\subsection{Exact solution for $N=Q=1$ in large vortex limit}
Assuming that the size $\xi$ of the vortex core in $M(r)$ is much larger than the size of the zero mode, which is set by inverse $M(\infty)\equiv M_\infty$, we may approximate\cite{Babak,IgorLu,Lu_JphysA} $M(r)=M_{\infty}r/\xi$. The resultant Hamiltonian reads
\begin{eqnarray}
  H 	&=& \gamma_0[ \alpha (\pi_x \gamma_1+\pi_y\gamma_2)+ M(r) \left(\cos\theta+ i\sin\theta\gamma_5\right)],\nonumber\\
  	&=& \Omega_c\left[\pi_x\Gamma_1+\pi_y\Gamma_2+\lambda(x\Gamma_3+y\Gamma_5)\right]\label{HH}
\end{eqnarray} 
where $\pi_i = p_i - A_i$ and, in the second equation, the parameters $\Omega_c=\sqrt{2|B|}\alpha$ and $\lambda={M_{\infty}}/({\Omega_c\xi\sqrt{|B|} })$ and the coordinates are rescaled as $r\to r \sqrt{|B|/2}$. The matrices $\Gamma_1 = \gamma_0\gamma_1, \Gamma_2 = \gamma_0\gamma_2, \Gamma_3 = \gamma_0$ and $\Gamma_5 = i\gamma_0\gamma_5$ form a closed Clifford algebra and each squares to identity matrix. The square of the Hamiltonian becomes a sum, 
\begin{equation}
  H^2=\Omega_c^2(\mathcal F_1+\mathcal F_2)\:.
\end{equation} 

The first term comes from the orbital part, which reads
\begin{equation}
  \mathcal F_1=\pi_x^2+\pi_y^2+\lambda^2(x^2+y^2)\:,
\end{equation}
In terms of the two operators
\begin{equation}
  a=\partial_{\bar x}+\frac{\bar x}{2}\:, b=\partial_{\bar y}+\frac{\bar y}{2}\:,
\end{equation}
where $\bar x=\sqrt{\mu}x$, $\mu=\sqrt{1+4\lambda^2}$ , we arrive at the following, 
\begin{equation}
  \mathcal F_1=\mu\left[a^{\dag}a+b^{\dag}b+1+\frac{i}{\mu}(a^{\dag}b-b^{\dag}a)\right]\:.
\end{equation}
Observing the following commutators,
\begin{equation}
  [\mathcal F_1,a^{\dag}]=\mu a^{\dag}-ib^{\dag}\:,
\end{equation}
and
\begin{equation}
  [\mathcal F_1,b^{\dag}]=\mu b^{\dag}+ia^{\dag}\:,
\end{equation}
it can be shown that the new operators $c^{\dag}_{\pm}=\frac{1}{\sqrt{2}}(a^{\dag}\mp ib^{\dag})$ diagonalize $\mathcal F_1$ such that $[\mathcal F_1,c^{\dag}_{\pm}]=(\mu\pm 1) c^{\dag}_{\pm}$. Consequently, the orbital contribution is expressed as,
\begin{equation}
  \mathcal F_1=(\mu+1)c_+^{\dag}c_++(\mu-1)c^{\dag}_-c_-+\mu\:,
\end{equation}
which has the lower bound of energy $\mu$.

The second term results from the commutators between coordinate and momentum operators,
\begin{equation}
  \mathcal F_2=\lambda[\pi_x,x]\Gamma_1\Gamma_3+\lambda[\pi_y,y]\Gamma_2\Gamma_5+[\pi_x,\pi_y]\Gamma_1\Gamma_2\label{F2_1}\:,
\end{equation} 
which in the matrix form reads,
\begin{equation}
  \left(\begin{array}{cccc}
  -\nu & 0 & 0 &0\\
  0 & \nu & -2i\lambda & 0\\
  0 & 2i\lambda & -\nu & 0\\
   0 & 0 & 0 & \nu
  \end{array}\right)\:,
\end{equation} 
where $\nu={\rm sgn}(B)$ is defined from the commutator $[\pi_x,\pi_y]$ in Eq.\ (\ref{F2_1}). In addition, it is useful to note that $\mathcal F_2^2=1+2\lambda^2-2\lambda^2\Gamma$, where $\Gamma=\gamma_0\gamma_3 = -\Gamma_1\Gamma_2\Gamma_3\Gamma_5$. So, that the eigenstates of $\mathcal{F}_2^2$ with eigenvalues 1 and $\mu^2$ are, respectively, eigenstates of $\Gamma$ with eigenvalue $\pm 1$. Therefore, $\mathcal F_2$ acting on the zero mode should give $-\mu$ to cancel the contribution of $\mu$ from $\mathcal F_1$. 

Hence the orbital part of zero mode $|\psi^0\rangle$ is found by,
\begin{equation}
  a|\psi^0\rangle=b|\psi^0\rangle=0\:,
\end{equation} 
from which one can see that $\langle \bar x,\bar y|\psi^0\rangle\sim \exp({-\frac{{\bar x}^2+{\bar y}^2}{4}})$ is also annihilated by both $c_{\pm}$. The spinor part of the zero mode is determined by,
\begin{eqnarray}
  \Gamma|\psi^0\rangle&=&-|\psi^0\rangle\:,\label{spinor1}\\
  \mathcal F_2|\psi^0\rangle&=&-\mu|\psi^0\rangle\label{spinor2}\:.
\end{eqnarray}
It is interesting to observe that Eq.\ (\ref{spinor1}) together with (\ref{spinor2}) act as projection onto the valleyspin-$\frac{1}{2}$ subspace supported on sublattice B. We may write $\mathcal P_+=(\id-\Gamma)/2$ as the projection onto sublattice B. The two-component valley spinor $(u,v)^T$ is then determined by the projected matrix $\tilde{\mathcal F_2}=\mathcal P_+\mathcal F_2\mathcal P_+=2\lambda\sigma_y+\nu\sigma_z$,
\begin{equation}
  \tilde{\mathcal F_2}\left(\begin{array}{cccc}
  u\\
  v\end{array}\right)=-\mu\left(\begin{array}{cccc}
  u\\
  v\end{array}\right)\:.\label{0mode_valleyspin}
\end{equation}
Now it is easy to show that the zero mode is,
\begin{equation}
  |\psi^0\rangle\propto
  \left(\begin{array}{cccc}
  0\\
  2i\lambda \\
  \nu+\mu\\
  0
  \end{array}\right)\exp{\left(-\frac18{\mu |B| r^2}\right)}\:.
\end{equation}
In the strong field limit, $\lambda\to 0$ and $\mu\to 1+2\lambda^2$, the valley-spin of zero mode is anti-aligned with the external field, i.e. $\langle\gamma_5\rangle=-\nu$ in the representation. On the other hand, in the zero field limit, $\mu\to2\lambda\gg 1$, the zero mode has vanishing valley-spin projection, $\langle\gamma_5\rangle=0$, which is expected as the time-reversal invariance is restored.

\end{document}